\documentstyle[aps,twocolumn]{revtex}
\def\be{\begin{equation}}
\def\ee{\end{equation}}
\begin{document}
\title{Random matrix approach to `nonuniversal' conductance}
\author{Petr \v Seba$^{1,2}$, Karol
\.Zyczkowski$^3$, and  Jakub Zakrzewski$^{3}$ }
\address{
$^1$ Nuclear Physics Institute, Czech Academy of Sciences, \\
250 68 Re{\v z} near Prague, Czech Republic }
\address{
$^2$ Pedagogical University, Hradec Kralove, Czech Republic }
\address{
$^3$Instytut Fizyki Mariana Smoluchowskiego, Uniwersytet  Jagiello\'nski,
\\ ul. Reymonta 4,  30-059 Krak\'ow, Poland}
\date{\today}
\maketitle
\begin{abstract}
Recent experiments on the conductance of high quality quantum wires
have revealed an unexpected feature: the quantization step of the
conductance is apparently system dependent.
 We provide the understanding of
this behaviour using the appropriately extended random matrix
approach. A single additional parameter  governs the size of
the conductance quantization steps. In effect the behaviour
seems to remain `universal', generic for the conductance of
a class of mesoscopic systems.

\end{abstract}

The conductance of mesoscopic devices, the so called quantum dots or
quantum wires, exhibits a number of universal
features such as the quantization of the average conductance or
the magnitude of the conductance fluctuations. For the ideal one-dimensional
(1D) quantum wire the dc conductance is quantized in units
of $G_0=2e^2/h$ (per channel in the wire, the factor 2 corresponding to
the electron spin) \cite{BeH91}, $G=G_0M_1$ with $M_1$ being
equal to the number of transverse modes supported by the wire.
Thus the dimensionless conductance $G/G_0$ changes by integer steps
when $M_1$ increases.
 Similarly, the average conductance of a quantum dot
coupled to the outside world by two leads each of which supports
$M_1$ open channels
is given by \cite{BM94}
\be
G=G_0\frac{M_1^2}{2M_1-1+2/\beta}
\label{mello}
\ee
where $\beta=1$ for time reversal invariant systems and $\beta=2$
when this symmetry is broken strongly. These results are readily
obtained within the random matrix theory (RMT) approach, recently
reviewed in detail by Beenakker \cite{Be97}.

Experiments, carried out recently on high quality wires \cite{yac96}
revealed quite surprisingly smaller quantization steps of the
height  $g<1$, with
$g$ varying from sample to sample and reaching 0.75 at low temperatures.
It has been pointed out \cite{yac96,ac97} that such a behaviour
may be an evidence of a coherent backscattering between the 1D wire
and the 2D leads. In such a case the conductance becomes $G=G_0T$
where $T$ is a $M_1$-dependent transmission coefficient.
 In the same work \cite{yac96} three different theoretical possibilities
 for the explanation of the data are discussed
The difficulties with the RMT approach and the Luttinger
liquid theory \cite{kane} are pointed out.
 The authors give their own explanation in terms of
the competition between the scattering from 2D into the edge modes.

The purpose of this communication is to show that the experimental
results may be, however, reproduced by the appropriate
 RMT model of scattering by a slight modification
 of the approach  which yielded in the past many successful
predictions for the transport properties in the mesoscopic media
\cite{Be97,lewd91}. The universal parameter determining the height of
the conductance steps is defined.

To this end let us assume that the almost ideal 1D wire is coupled
to 2-dimensional (2D) leads as realized in the recent experiment
\cite{yac96}. We consider a standard Heidelberg scattering matrix
approach \cite{mawd69} expressing the scattering matrix $S$ as
\be
S=1-2\pi i W^\dagger(E_F-H+i\pi W W^\dagger)^{-1} W,
\label{s}
\ee
where $H$ is the internal Hamiltonian of the system
represented by a matrix of rank $N$ while $W$ is a $N\times M$
matrix representing the coupling  between the $N$ internal
states and $M$ scattering channels in the leads. Assuming
two identical leads one gets $M=2M_1$.

In the application to a chaotic cavity scattering one assumes
that the number of internal states, $N$, around the Fermi energy, $E_F$,
is much larger than $M$. Taking typical RMT assumptions about the
statistical properties of $H$ and $W$ one may then derive a number
of predictions concerning the statistical properties of $S$ and
of the measurable observables. As shown by Brouwer \cite{bro95}, such an
approach is equivalent (for $M\ll N$), to making RMT assumptions
concerning directly the unitary $S$ matrix itself. For example,
if $H$ pertains to the Gaussian Orthogonal Ensemble (GOE)
 and $W$ are composed of real random
vectors (the situation appropriate for time reversal invariant systems),
 then the $M\times M$ matrix $S$ belongs to the corresponding
circular orthogonal ensemble of unitary matrices (COE) in the limit
$N\rightarrow \infty$. Similarly, if time reversal symmetry is broken
and $H$ pertains to Gaussian Unitary Ensemble (GUE), the corresponding
$S$ matrix shows statistical properties typical for the Circular
Unitary Ensemble (CUE).
 Thus it is justifiable to derive transport
properties by making statistical predictions for $S$ matrices
themselves. Such an approach yields, e.g., Eq.(\ref{mello}).
The advantage of the former, Heidelberg approach
 is that it allows
also to calculate energy dependent quantities such as correlators
or time delays,  while the  direct RMT approach to $S$ matrices says
nothing about the dependence on the scattering energy, $E_F$.

Consider now the experimental system of \cite{yac96}. The 1D almost
ideal wire placed between 2D leads takes the place of the internal
scattering system in the Heidelberg approach with $N$ being now
the number of states in the internal wire around $E_F$ or the number
`internal channels'.
Note that really the Hamiltonian describing the internal wire supports an
infinite number of states. Most of them does not contribute to conductance
being  vanishingly small (evanecent)
 on the left or right side of the 1D wire. The
important $N$ ``states'' are the $N$ scattering channels through the 1D
wire if it were coupled incoherently to leads.
Thus $N$ can be even only.
Moreover there is no ground to assume that the internal matrix $H$
pertains to GOE. Rather, since the wire is almost ideal, one
may assume that the motion of the electrons through the wire is
ballistic.
Since the leads
are assumed to be two dimensional,
 $M=2M_1$ should be much {\it larger} than $N$.
Note that the limit $N\ll M$ is the opposite to that taken in the
standard transport theory \cite{Be97}.

The structure of the $S$ matrix, Eq.(\ref{s}), indicates
that $N$ out of $M$  of its eigenphases may be nontrivial and different
from 0 (i.e. the remaining $M-N$ eigenvalues of $S$ are equal to unity).
 This is due to the fact that the part coupling the channels to
the internal states has at most the rank $N$. Writing $S$ as
\be
S=\left(
\begin{array}{cc}
r & t\\
t' & r'
\end{array}
\right)
\label{mat}
\ee
one realizes that the dimensionless conductance
 $G/G_0={\rm Tr} tt^\dagger$
(the Landauer formula)
may depend only on  the parameter $c=N/M$. Further we
shall expect that the average conductance increases in steps
when $N$ changes. The size of the steps  may
depend on $c$.

To test this qualitative picture we have simulated the
conductance of the system by averaging the transmission obtained over
several random realizations of $H$ and $W$. In all the simulations
$E_F=0$ while we have varied $N$, $M$, as well as assumed different
statistical properties of $H$.
Specifically, we shall assume either GOE case or the situation when
the eigenvalues of $H$ are uncorrelated. The latter case we shall
call the Poissonian ensemble (PE) since the nearest neigbour
statistics takes then a Poissonian form.
 The  $N\times M$ coupling matrix $W$
was composed of $N$ mutually orthonormal random vectors of length
$M$.
The average is obtained by taking 1000
different realizations of a given system.
Fig.~\ref{fig1}(a) shows the average transmission (dimensionless
conductance) obtained keeping a fixed value of $c=N/M$ and increasing
$N$ by two. Observe that regardless of the properties of the internal matrix $H$
 the qualitative behaviour of the conductance is quite
similar, it increases in steps smaller than unity, the value of the
step being dependent on $c$ and to a much lesser extend on the
statistical properties of $H$.
 Panel (b) shows the behaviour of the system
while keeping fixed the number of `internal channels' $N$ and
increasing $M$. Observe that the conductance steps actually {\it decrease}
with $M$ for $M$ large. It is the number of `internal channels', $N$,
which limits the conductance value. The dependence on $M$ is
much weaker and indicates that
for larger $M$ the backscattering
 plays a larger role leading to decrease of the conductance
steps.

Note that it is the backscattering on the abrupt transition between
the 1D wire and the 2D leads that is solely responsible for
the size of the conductance steps. Were the transition from wire
to leads a smooth one, no backscattering would occur and the
conductance steps would be equal to unity as follows from
Levinson study \cite{levi}.

Fig.~\ref{fig1} shows already that the experimental observations
of \cite{yac96} may be at least qualitatively explained by the simple
RMT model. To exemplify this point further we have assumed that
the density of states changes according to a triangular potential
well (as in the experiment) when the applied voltage is varied.
After choosing the free parameter in the model, i.e., $c$, the
conductance dependence on the applied voltage reproduces fairly
accurately the Fig.~2 of \cite{yac96} (see Fig.~\ref{ourexp}).

Let us point out that the results obtained are very weakly
dependent on the statistical properties of the internal
Hamiltonian $H$. For a given $c$, the conductance quantization step,
observed when $N$ is varied, increases slightly as the statistical
properties of $H$ change from PE to GOE or
the picket fence spectrum corresponding to the levels repulsion
parameter $\beta\to \infty$. The quantization step size remains
practically unaffected (within the statistical significance of our data)
if we consider the case of broken time reversal symmetry, i.e.,
with $H$ belonging to GUE.

Our simple model cannot account for the changes of the conductance
steps with the temperature, $T$ (as observed in \cite{yac96} for
larger $T$ the conductance step size increases and approaches unity).
Such temperature changes are indicators of the importance of the
electron-electron (e-e)  interactions \cite{yac96}.
 It seems thus quite intuitive to
blaim this interaction also for the non-integer conductance steps.
In this respect the fact that
our model, being a  single particle approach,
also yields $c$ dependent conductance steps is quite surprizing.
Apparently, the step size can be reconstructed from RMT, i.e. a single
particle approach (where at least a part of e-e interaction may be in
principle included via Hartree or mean field approach).
The increase of the step size with temperature will then point out to
the increasing importance of the genuine many particle effects.

In effect, our model
is appropriate, strictly speaking, for low temperatures only. Still
qualitatively, one may explain the increase of the conductance
with $T$ in terms of decreasing backscattering as mentioned in
\cite{yac96}. The point is that when temperature increases, the internal Hamiltonian
$H$, described for low temperatures by PE, has to be - due
to the increasing role of the temperature dependent disorder -
replaced by a GOE matrix, and this leads to an increase
in the conductance step.


Accepting that the model presented yields reasonable predictions
concerning the conductance steps one can ask whether in the
studied, $N\ll M$ case  similar predictions may be obtained
using RMT assumptions directly for the $S$ matrix. Naturally,
the standard approach \cite{Be97} has to be modified since the
$S$ matrix, must have $M-N$ unity eigenvalues.

We are thus going to mimic the scattering matrix by  $S=UDU^{\dagger}$,
where $D$ is a diagonal matrix consisting of $M-N$ unit eigenvalues
and $N$
 eigenvalues  $\exp(i \varphi_i)$. The nontrivial eigenphases
 $\varphi_i$ are distributed according
to some assumed joint probability distribution
 $P_{\beta}(\varphi _{1,\cdots ,}\varphi _N)$, characterized by the
level repulsion parameter $\beta \in [0,\infty]$. Random unitary
rotation matrix $U$ is drawn uniformly with respect to the
Haar measure on $M$ dimensional unitary space and pertains to CUE.
Such an assumption concerning $U$ is fully coffect for a  broken
 time-reversal
invariance, the situation not realized in the experiment \cite{yac96}.
We know, however, from the standard RMT of scattering (in the $M\ll N$
limit) that the dependence of the conductance on the symmetry  is
relatively small for  disordered wires and appears only on the level
of weak localization corrections \cite{Be97}
 through the eigenphases repulsion
parameter $\beta$. Thus the results obtained should only weakly
depend on detailed properties of $U$. This assumption is even more
justified by the numerical results, mentioned above, that revealed
that the conductance step size is not sensitive to the change from GOE
to GUE within the Heidelberg model.

 The total  conductivity in the system is given
 by a sum of the individual transmission
coefficients $G/G_0=\sum_{l=M/2+1}^M\sum_{m=1}^{M/2}|S_{lm}|^2$.
There exist $M^2/4$ elements of the matrix $S$,
 contributing to the
total conductance. Each element of this sum can be written as
 $S_{lm}= \sum_{k=1}^NU_{kl}^{*}U_{km}(e^{i\varphi _k}-1)$,
with $l\ne m$. The double average
 $\left\langle \left\langle |S_{lm}|^2
 \right\rangle _U\right\rangle _D $,
 over $N$ random phases of the diagonal matrix $D$
and over random rotation matrix $U$,
consists of $N$ diagonal and $N(N-1)$ off-diagonal terms.
 The averages over
unitary matrices $U$ distributed according to the Haar measure are
known \cite{mello90}. They allow us
to compute the average conductance at least for the two limiting cases,
namely uncorrelated eigenphases
 ($\beta=0$ - Poissonian case), and picket fence ($\beta\to
\infty$ - equally spaced) distribution. We get

\begin{equation}
(G/G_0)_{P}=
 \frac{MN}{4(M^2-1)} \left( 2M-N-1\right).
\label{PPP}
\end{equation}
for the Poissonian case and
\begin{equation}
(G/G_0)_{C}=
\frac{MN}{4\left( M^2-1\right) }\left( 2M-N\right),
\end{equation}
for the most rigid cristalline, picket fence case. Clearly, the
results for other ensembles should lie between these two
limits.

Using  the above formulae one may calculate the conductance steps
for fixed $c=N/M$. In the limit of large $N$, the step $g$
(when $N$ increases by two) is equal to $g_P=1-1/c$ for the
former and $g_C=1-1/2c$ for the latter distribution.
Clearly, also the model constructed to mimic the $S$ matrix directly
is capable to yield the prediction for the conductance step size
smaller than unity.

While both the approaches, the Heidelberg method and the direct
modelling of the $S$ matrix properties yield similar predictions
for $N\ll M$, i.e. give conductance steps smaller than unity,
the models seem not to be equivalent (as in the opposite case
of $M\ll N$ - see \cite{bro95}). For example, assuming in the
former approach that the internal $H$ pertains to GOE does not
assure that the nontrivial eigenphases of the $S$ matrix obey the
appropriate COE statistics (as has been checked numerically).
Still, as shown in Fig.~\ref{comp}, for $N\ll M$ and fixed
$c$ both models yield quite similar prediction for the average
conductance (and thus the size of the quantization steps).
This robustness of the nonuniversal step size to the details
of the random model assumed suggests strongly that the
phenomenon is quite general and occurs whenever the number of
open channels $M$ exceeds the number of internal states.

Apart from the pure wire case discussed in \cite{yac96} one can
envision a chaotic quantum dot (with many thousands of levels)
coupled by two almost ideal 1D leads to the broad connectors. Provided
the coherence length exceeds the length of 1D leads we expect
coherent backscattering on the border between the leads and the
connectors. Then the number of original channels in the 1D leads
determines the number of extended states, $N$, in the system: quantum dot +
leads. This number may be quite small. All other levels of the
quantum dot remain localized and do not contribute to the conductance.
As the number of channels in the connectors, $M$, is large, the
situation $N\ll M$ is recovered and we predict that the conductance
quantization steps will be smaller than unity and will depend on
the ratio $c=N/M$.

To summarize, we have presented simple, Random Matrix Theory
based models, which predict that the conductance steps may be
smaller than unity in agreement with the recent experiment
\cite{yac96}. The origin of the deviation from unit steps comes
from the coherent backscattering on the border between interior
and the exterior of the system. The effect is quite general,
does not depend on the details of the model, the only requirement
being that the number of internal states, $N$, is much
 less than the number of open channels, $M$.
Since the RMT model is essentially a single particle model
our results indicate that electron-electron interactions can
not be a sole origin of the ``nonuniversal'' conductance steps.

 This work was  supported by
the Polish Committee of
Scientific Research under grant
No.~2P03B~03810 (K.\.Z. and J.Z.) and by the
Theoretical Physics Foundation in Slemeno
(P.\v S.).
\newpage


\begin{figure}
\caption{Panel (a) displays mean dimensionless conductance $G/G_0$ evaluated
for c=0.5 (thick lines) and 0.3 (thin lines).
 The full line corresponds to the Poissonian case, broken
lines represent results obtained for GOE.
Panel (b) shows $G/G_0$ as a function of the number of
channels M. The number of internal states $N=20$.
Thick (thin) line corresponds to  GOE  (PE) case,
respectively.
}
\label{fig1}
\end{figure}

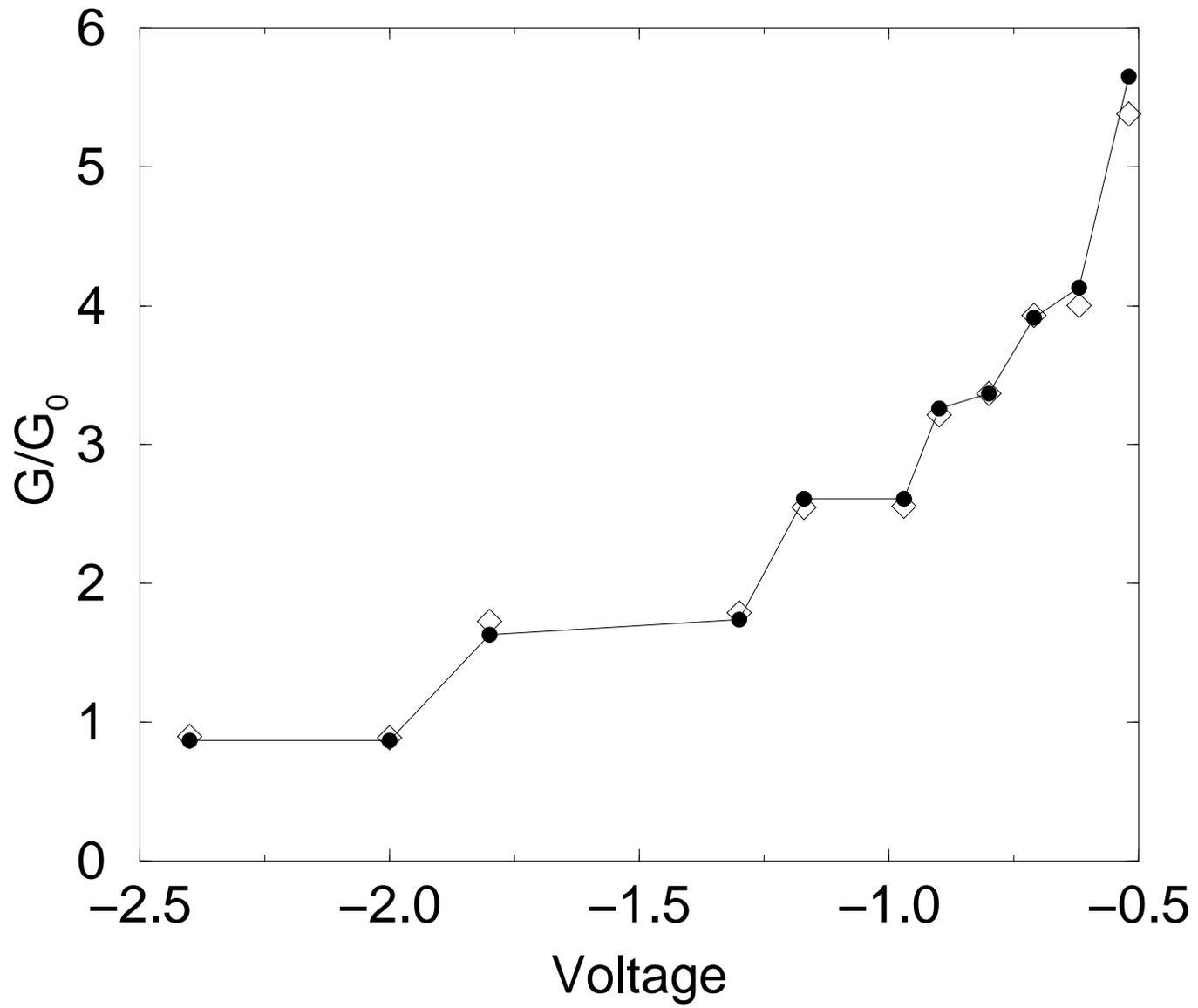
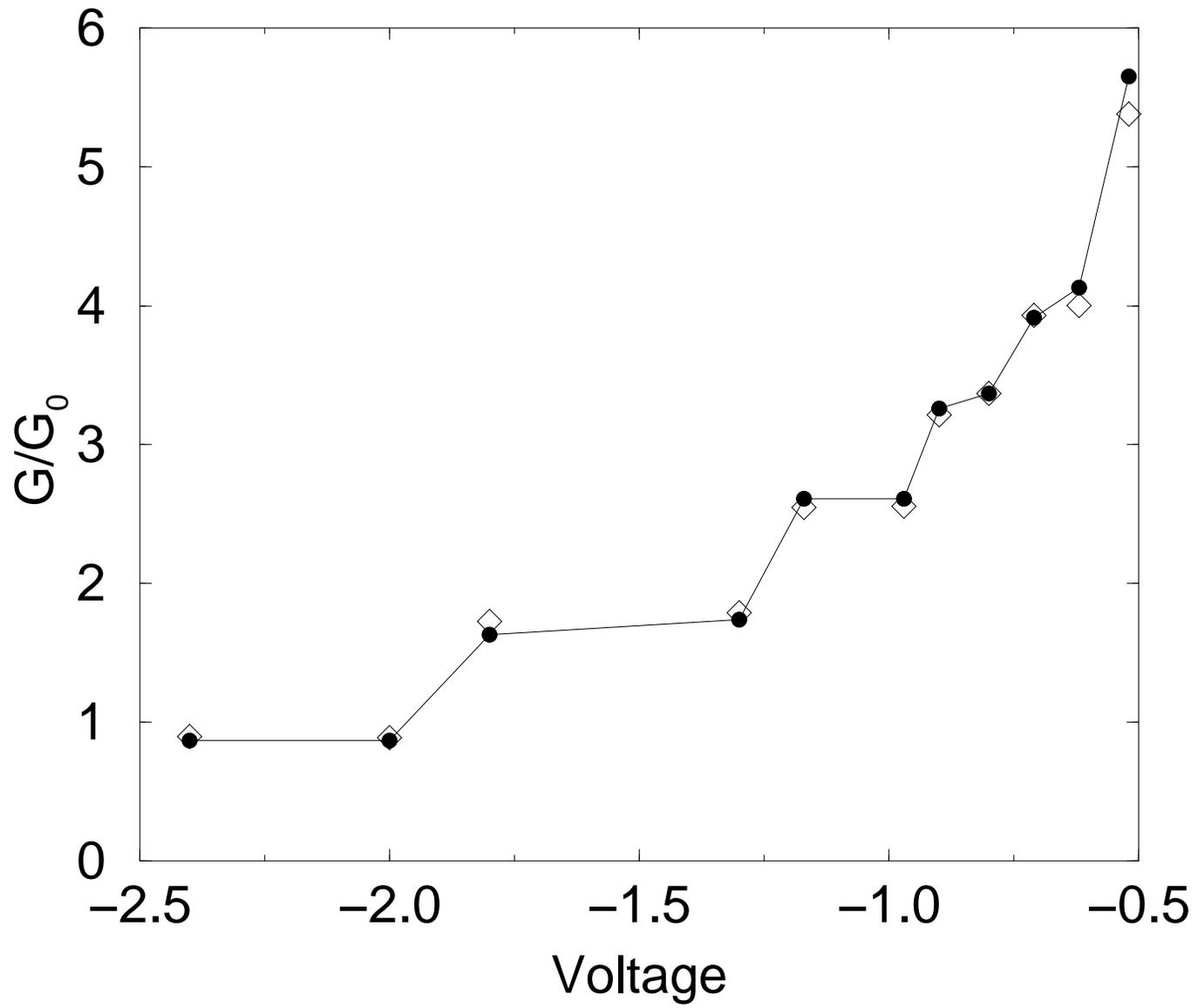
\begin{figure}
\caption{Dependence of the dimensionless mean
conductance on the applied voltage $V$. Filled dots connected by
the line (to guide the eye) represent the results
 obtained in our model calculations
with Poissonian internal matrix. The number of the channels $M$ and
internal states
$N$ depends on the voltage $V$ as 
$M=[[(a-3.7/V)]]$; $N=[[-3.7/V]]$ with $a=8.8$.
Here $[[x]]$ represents  even number being most close to $x$. Diamonds
correspond to the
experimental results obtained in [4].  }      
\label{ourexp}
\end{figure}

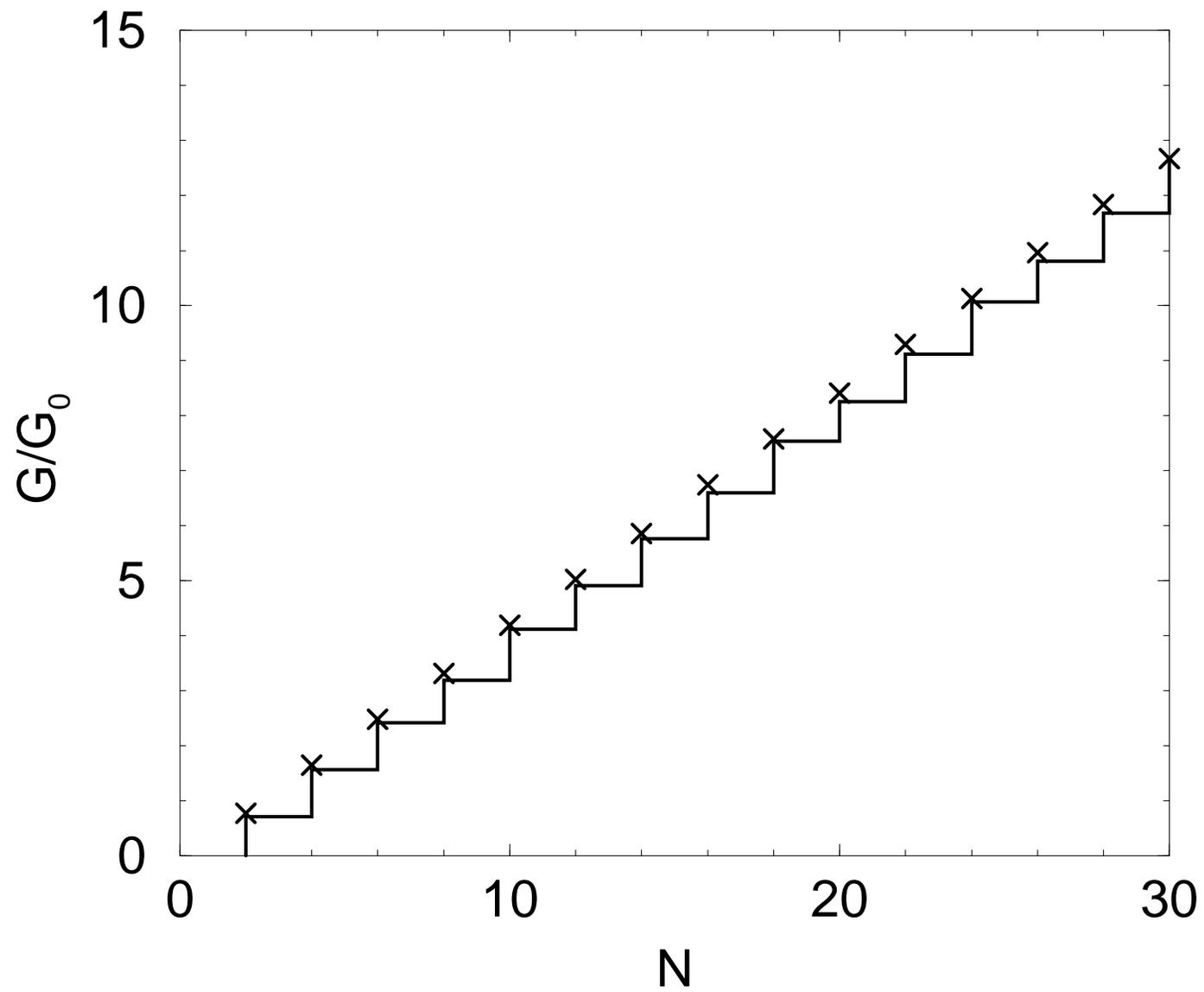
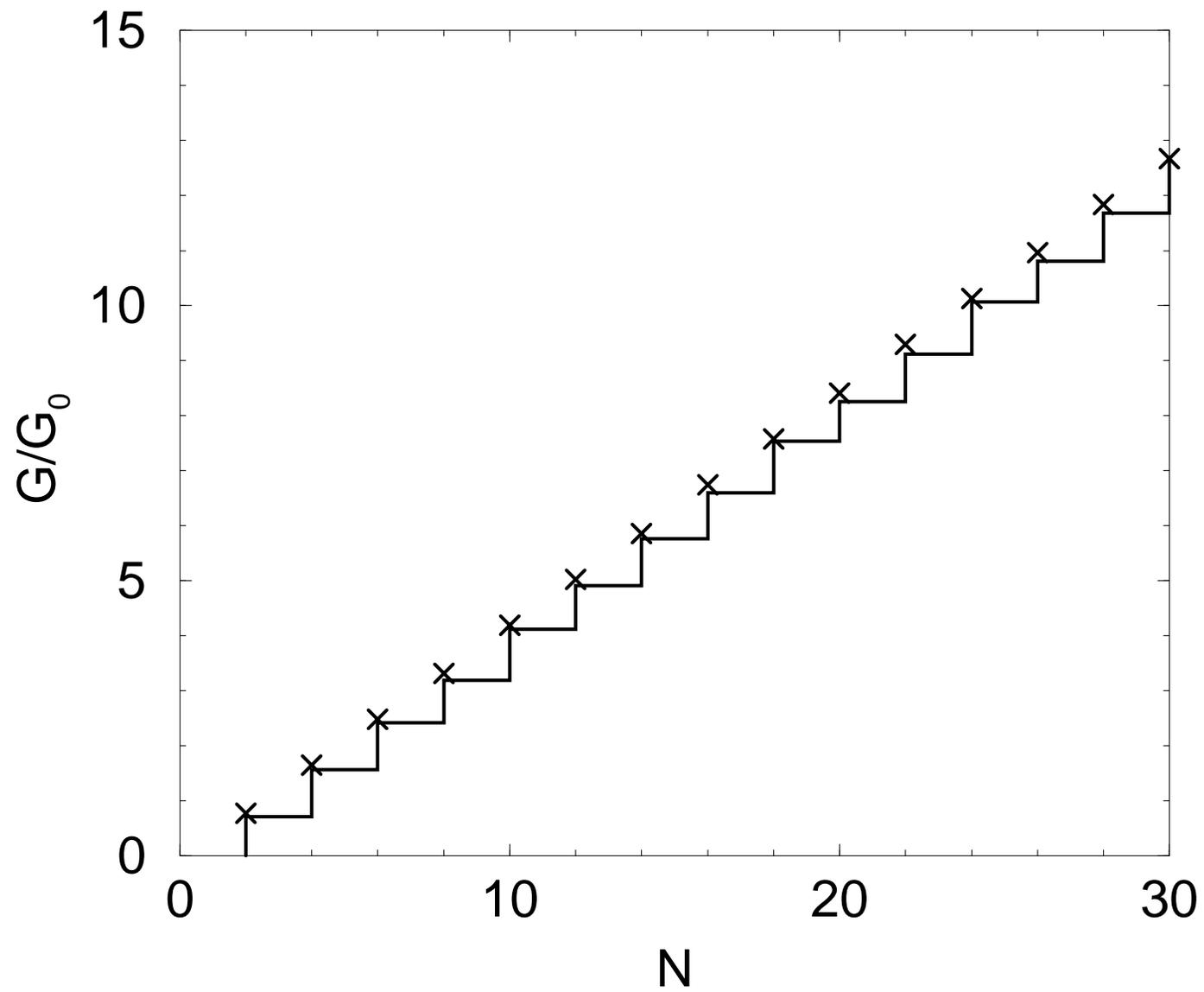
\begin{figure}
\caption{
Mean conductance obtained for c=0.3 in the Poissonian case is
ploted as a function of N (full line) and compared with the
Eq.~(\protect{\ref{PPP}}) (crosses).  }
\label{comp}
\end{figure}

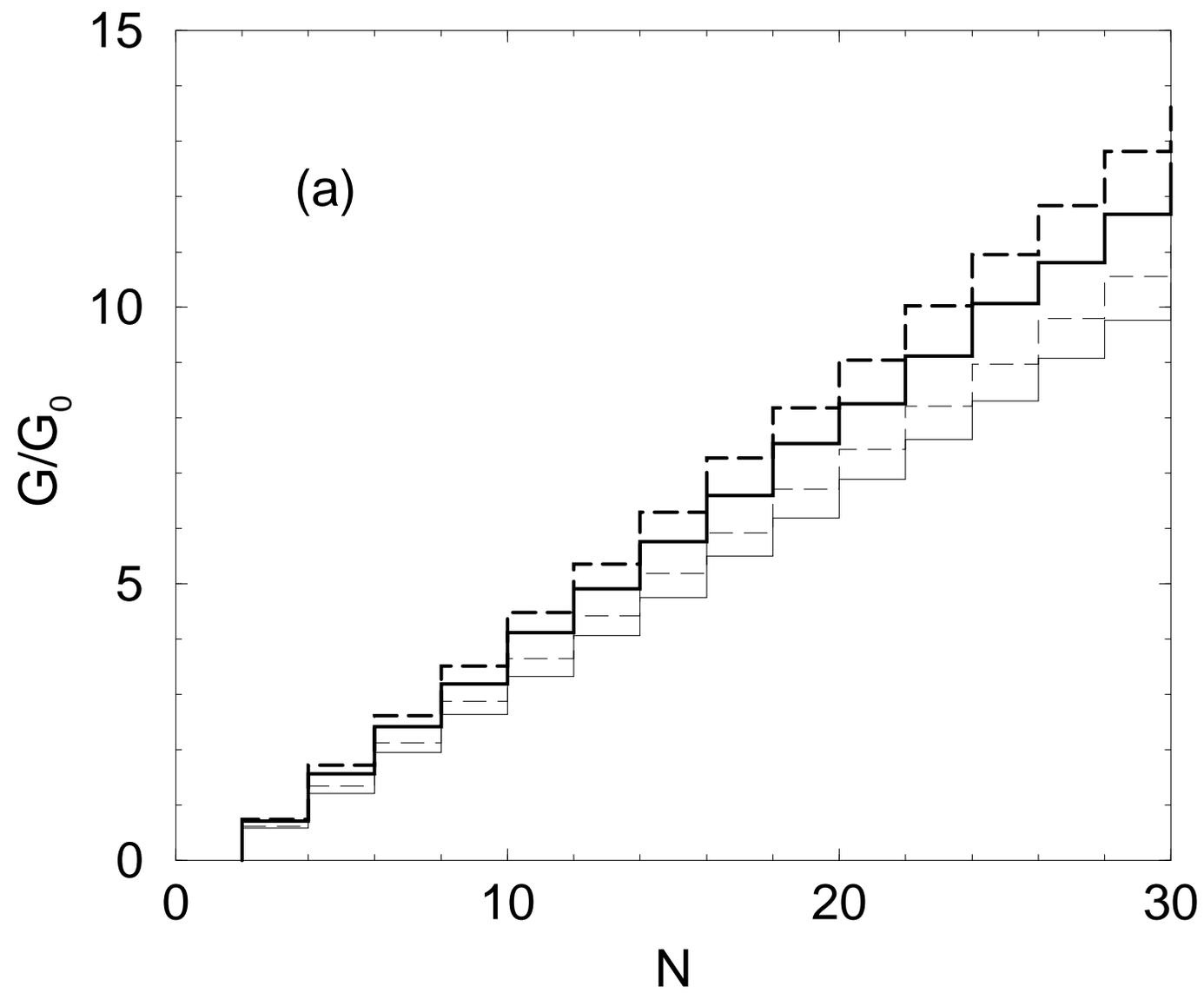
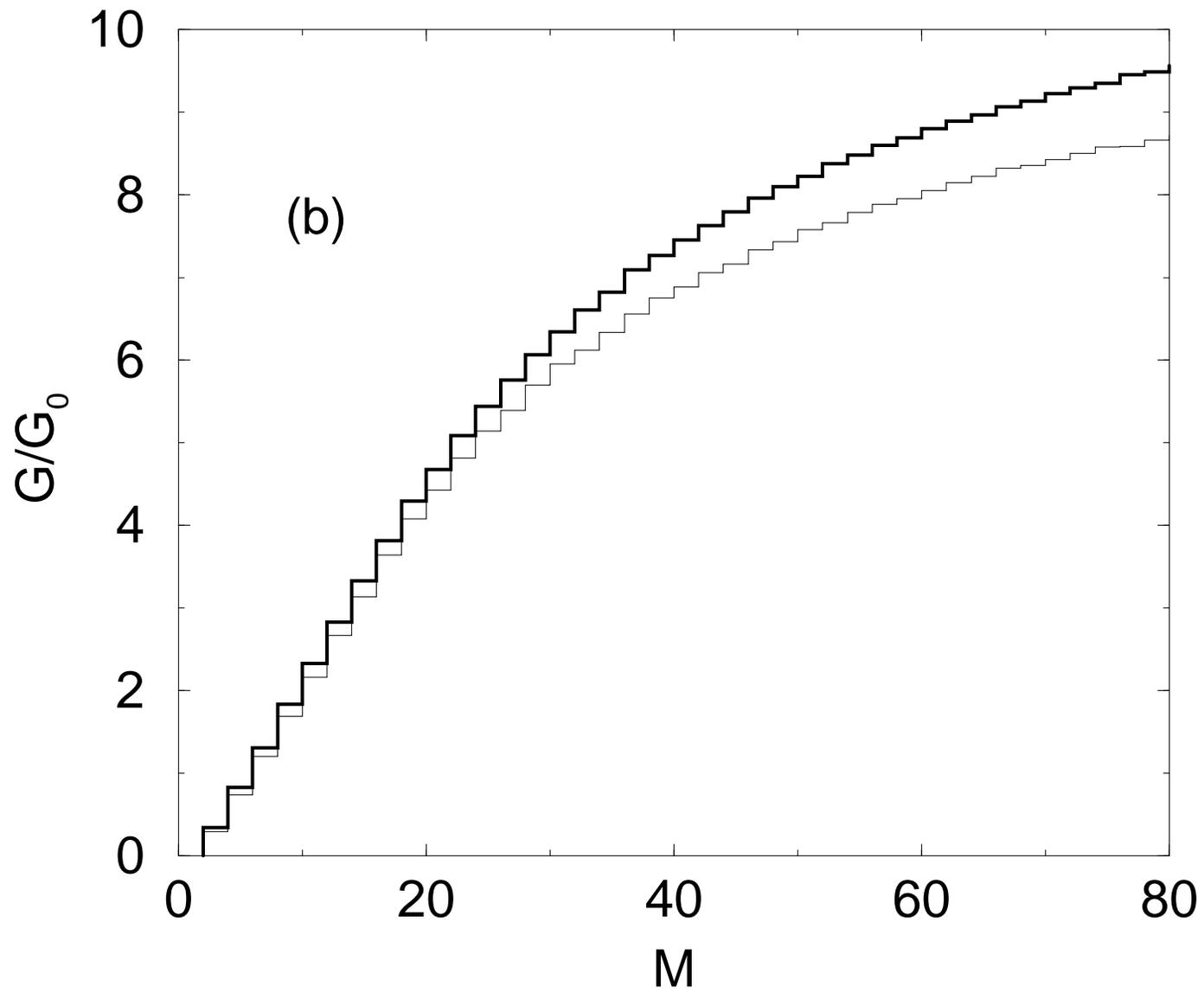
\end{document}